\documentclass[11pt]{article}
\usepackage{amsmath, amsthm, amssymb}    
\usepackage{bridges}			
\usepackage{graphicx}			
\usepackage[colorlinks=true, urlcolor=blue, citecolor=black, linkcolor=black]{hyperref}  
\usepackage{subcaption}			

\usepackage{color}
\graphicspath{{./figures/}{./dsa_material/}}
\usepackage{subcaption}
\usepackage[format=plain, justification=justified, indention=1cm]{caption}

\newcommand*{\kdots}{\dots\unkern}

\title{Vector Fields and Paul Klee -- A Summer School Course\\ for gifted High-School Students}

\author{Martin Skrodzki\textsuperscript{1} and Henriette-Sophie Lipsch\"utz\textsuperscript{2}
\vspace{10pt}\\
\textsuperscript{1}RIKEN iTHEMS, Wako, Japan; Martin.Skrodzki@riken.jp\\
\textsuperscript{2}Freie Universit\"at, Berlin, Germany; Henriette.Lipschuetz@fu-berlin.de} 

\date{}					

\begin{document}

\maketitle

\thispagestyle{empty}

\begin{abstract}
The artist Paul Klee describes and analyzes his use of symbols and colors in his book ``P\"adagogisches Skizzenbuch'' (pedagogical sketchbook, 1925). 
He uses arrows for means of illustrations and also discusses the arrow itself as an element of his repertoire of symbols.
Interestingly, his point of view on arrows matches the way in which vector fields are represented graphically.
Based on this connection of mathematics and Klee's artwork, we developed a concept for a summer school course for gifted high-school students.
Its aim was to both introduce the participants to higher-level mathematics and to let them create their own piece of artwork related to the art of Paul Klee and vector fields.
This article elaborates on the connection between Klee's art and mathematics.
It furthermore presents details of the summer course and an evaluation of its goals, including resulting artwork from participating students.
\end{abstract}

\section*{Paul Klee and Arrows}

Paul Klee (1879--1940) was one of the most influential and well-known artists of the Bauhaus movement. 
He taught at the ``Staatliches Bauhaus'' in Weimar, Germany, from 1920 until 1931.
During this time, he wrote 
 the ``P\"adagogisches Skizzenbuch'' (pedagogical sketchbook\footnote{All translations are by the authors. For longer texts, the original is omitted and translated passages are indicated by $^\mho$.}, \cite{Klee1925paedagogisches}).
These works enable us to interpret Klee's painting on the foundation of his own explanatory writings.

\begin{figure}
	\centering
	\begin{subfigure}[t]{0.45\textwidth}
		\centering
		\includegraphics[width=0.65\textwidth]{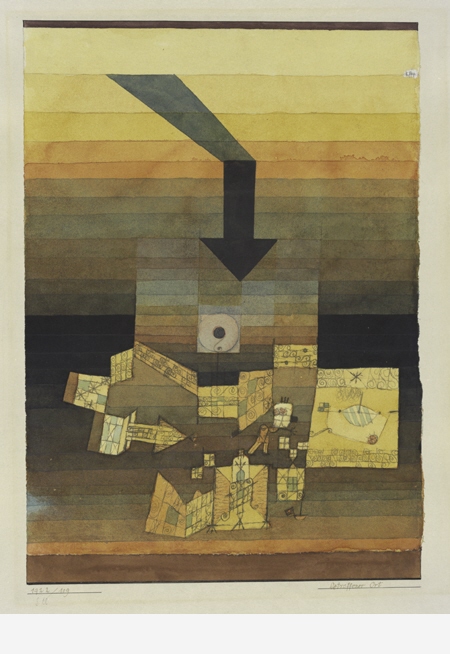}
		\caption{``Betroffener Ort'', 1922.}	
		\label{fig:BetroffenerOrt}
	\end{subfigure}
	\hfil
	\begin{subfigure}[t]{0.41\textwidth}
		\centering
		\includegraphics[width=0.55\textwidth]{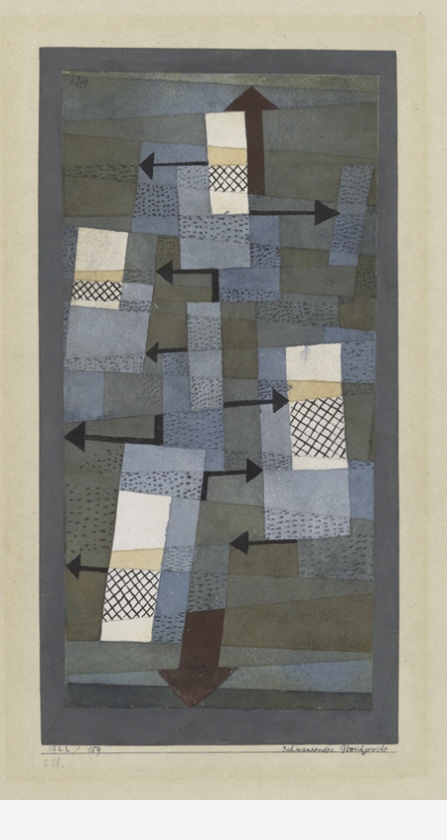}
		\caption{``Schwankendes Gleichgewicht'', 1922.}	
		\label{fig:SchwankendesGleichgewicht}
	\end{subfigure}
	\caption{Two works of Paul Klee featuring arrows. Courtesy of the Zentrum Paul Klee, Bern.}
\end{figure}

Several works of Paul Klee contain arrows in different forms and quantities. 
For instance, the painting ``Schwankendes Gleichgewicht''~(wavering balance, 1922) shows a multitude of arrows, see Figure~\ref{fig:SchwankendesGleichgewicht}.
On the other hand, ``Betroffener Ort'' (affected place, 1922) shows only two arrows, see Figure~\ref{fig:BetroffenerOrt}.
Both works feature straight and bent arrows, where the latter provide an increased dynamic to the paintings.
The first painting of Paul Klee to feature the word ``Pfeil'' (arrow) in its title is ``Landschaft mif fliegenden V\"ogeln einer von dem Pfeil durchbohrt'' (landscape with flying birds, one pierced by an arrow, 1918). Several more works feature arrows directly in the title or as symbols utilized in the painting, see~\cite{Paul2004Catalogue}.

The significance of the arrow in Paul Klee's work is most notable in his own writings.
In his pedagogical sketchbook~\cite{Klee1925paedagogisches}, the sections 37 to 40 out of a total of 43 sections are devoted directly to the arrow. 
Klee distinguishes the ``real'' and the ``symbolic'' arrow~\cite[Section 38, p. 44]{Klee1925paedagogisches}.
He characterizes the arrow as ``the father of the thought: How do I extend my reach there?''$^\mho$.
In the subsequent parts of his sketchbook, Klee continues to explain that a color gradient is related to a certain sense of direction and orientation~\cite[Sections~40\&41, pp.~47--49]{Klee1925paedagogisches}.
This latter sense exactly matches the use of colors for example in topographic maps, where they indicate the gradient direction of the scalar height field. 
Furthermore, Klee describes the arrow as a ``synthesis of cause and effect''$^\mho$, 
thus giving it not only a spatial, but also a temporal directedness.
For a more detailed discussion on the use of arrows in Paul Klee's work, we refer the reader to~\cite{Moesser1977pfeile}.

Klee himself was interested in the natural sciences. 
In a talk on Paul Klee's notebooks, researcher Andrzej Herczynski summarizes that ``[Paul Klee's] notebooks are a trove of geometrical and physical ideas aiming at a synthesis of natural and mathematical form.''~\cite{Herczynski2017Paul}.
This is evident for example in the title of Klee's painting ``Mathematisches Bildnis'' (mathematical likeness, 1919).
These connections and statements by Klee himself build the foundation of a course concept connecting mathematics and Klee's artwork.
We developed such a course for a summer school for gifted high-school students.
In the following, we will present this course's content and framework.

\begin{figure}[b]
	\centering
	\begin{subfigure}{0.45\textwidth}
		\includegraphics[width=1.\textwidth]{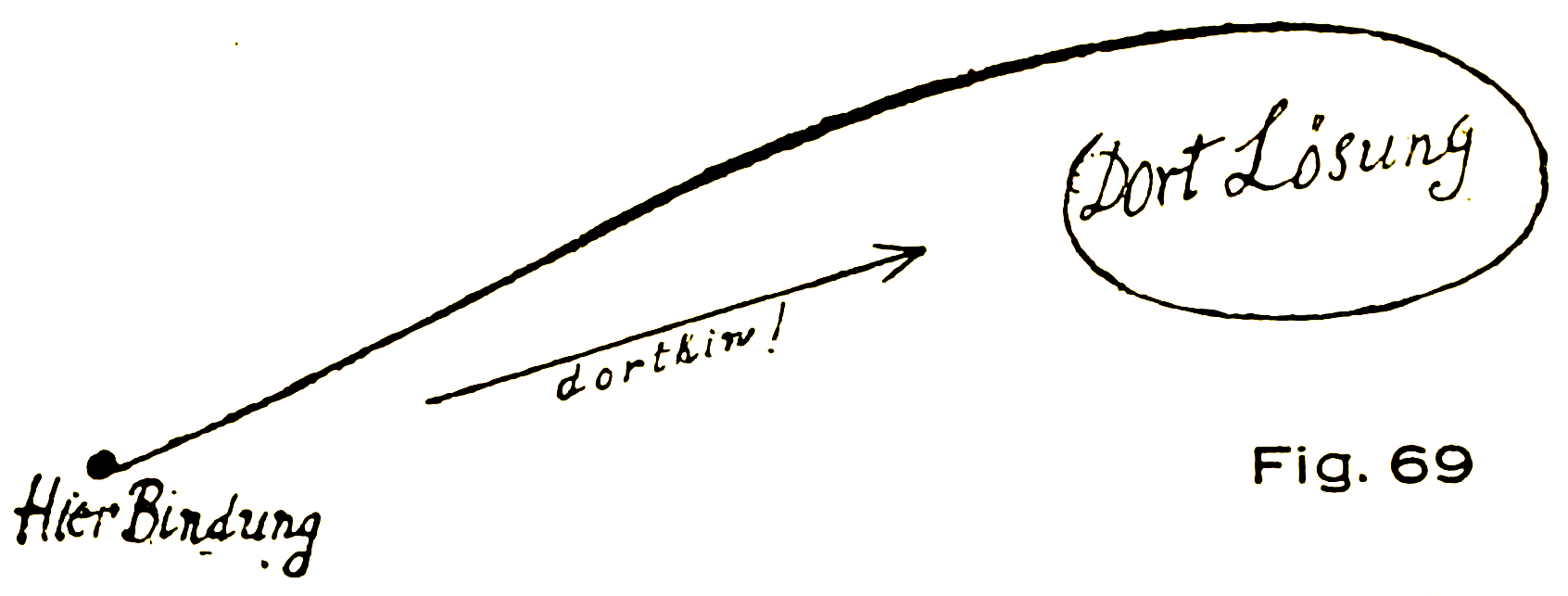}
		\caption{Figure 69, \cite[p. 44]{Klee1925paedagogisches}. This figure is placed to illustrate the characterization of the arrow. It reads (from left to right): ``connection here'', ``to that place!'', and ``solution there''.}
		\label{fig:Sketchbook69}
	\end{subfigure}
	\hfill
	\begin{subfigure}{0.45\textwidth}
		\includegraphics[width=1.\textwidth]{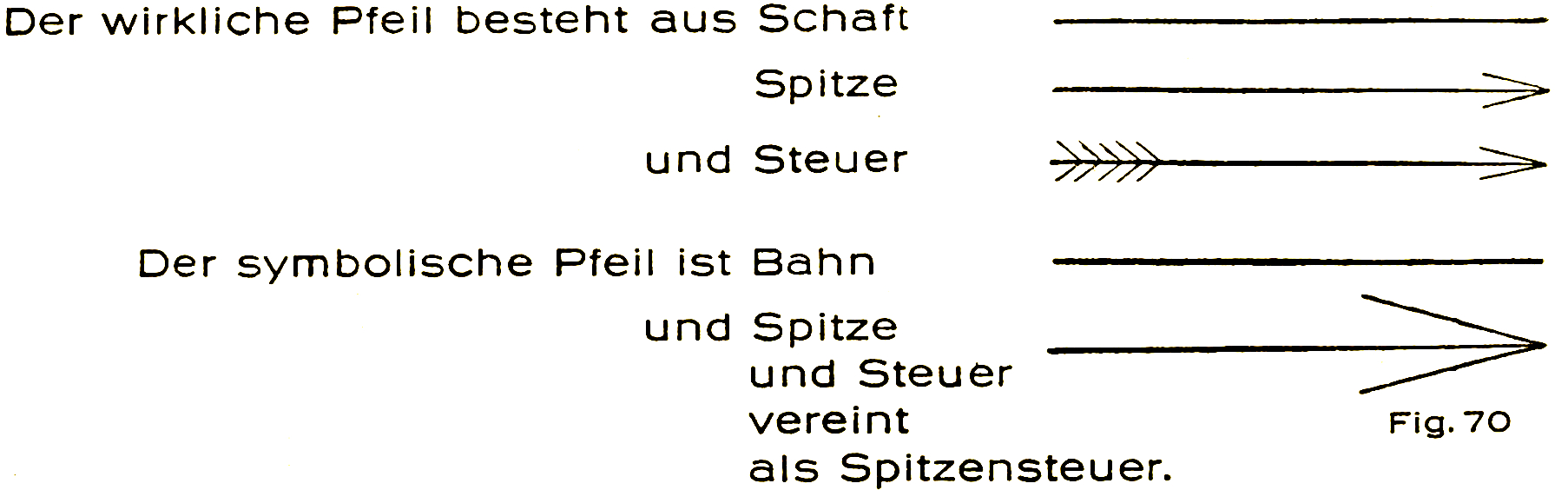}
		\caption{Figure 70, \cite[p. 44]{Klee1925paedagogisches}. This figure highlights the difference between the ``real'' and the ''symbolic'' arrow. It reads: ``The real arrow has a shaft, tip, and fletching. The symbolic arrow is track and tip and fletching integrated as tip-fletching.''}
		\label{fig:Sketchbook70}
	\end{subfigure}
	\caption{Figures from Klee's pedagogical sketchbook \cite{Klee1925paedagogisches}, illustrating his use and interpretation of arrows.}
\end{figure}

\vspace{-0.5em}
\section*{Content of the Course}

Arrows also play an important part 
 in mathematics. In high-school, they represent maps and vectors. In higher mathematics, they appear in many different contexts such as commutative diagrams or graph theory. In everyday life, colors and arrows depict for example temperature and wind direction on a weather map.

The course revolved around mathematics, art, and their relations. Basic knowledge of vector and differential calculus was required for participation. Building on it, the participants received a collection of notes on first year university vector calculus and linear algebra.
 Every student independently prepared a talk on a given topic and presented it during the summer school. While the students had to cope with the scientific level of notation and precision comparable to that at a university, a lot of examples were considered to illustrate the recently learned definitions. 

Starting with notions of sequences and convergence, the necessary basics in calculus such as continuity and differentiability were repeated. Afterwards, this knowledge was transferred to higher dimensional situations. The students came in touch with objects like vector fields and operations on them, such as gradient, curl, or divergence. As instructors, we focused on a high level of scientific quality in the talks as well as in the additional explanations provided by us. There was no simplification of the theory, which was at first very challenging for the participants. In particular, the proper use of mathematical notation turned out to be problematic for them.


Aside from the mathematical structures introduced, in two further talks, both Paul Klee's biography and his pedagogical sketchbook~\cite{Klee1925paedagogisches} were introduced. To provide the students with ideas as to how mathematics and art can interact, two talks focused on publications from Bridges~\cite{Poelke2014Complex,Skrodzki2016Chladni} and a brief introduction to fractals was given by the authors. These served as inspiration for the artistic projects of the participants, see below.

\section*{Motivation and Course Framework}

In German high-schools, mathematics and arts are two distinct disciplines that are usually not brought into contact with each other. Inspired by the work done in the Bridges community, we aimed for a project to show high-school students possible connections of mathematics and arts. This is to overcome established ways of thinking and to show connections between topics that are not related at first sight. Towards this end, Paul Klee and his writings on his art are a reasonable choice for this project since both can be related to the mathematical theory listed above. Since the artist understood his work in terms of depicting movement and as development of directions and dimensions, his art can be analyzed following this guideline and can be understood mathematically in this sense.

Another motivation of the course was to provide an atmosphere, in which artistically inclined participants can gain insights into mathematical content and in which mathematically inclined students can experiment with artistic practices. As course leaders, it was particularly important for us to provide all participants with a positive experience in the mathematical parts to encourage them in the pursuit of a career in mathematics or related subjects.

The course took place in summer 2019 as part of the ``Deutsche Sch\"ulerAkademie'' (German pupil academy, DSA). Since 1988, the DSA has been the main institution for supporting gifted high-school students; it runs up to seven summer schools per year. Each of these has a period of thirteen days which provide 33 time slots (90 minutes each) for course work. Generally, the schools take part during the German summer break. Each school has six separate courses, representing different disciplines, and enrolling fifteen to sixteen participants---supervised by two tutors---per course. The attendees have to attend a German school,
 have to be in their 10th to 12th grade, cannot be older than 20, and have to have outstanding intellectual abilities. 
 The participants apply for up to three of the courses available at any of the schools. They are then matched to attend one of the courses they applied for. Every participant can only attend one DSA in their lifetime. The courses are taught by volunteers with diverse backgrounds, selected based on their qualifications and the quality of their project proposals. For further information, see~\cite{DSA-Webpage} (only German).

Our course had 16 participants (10 female, 6 male) aged from 15 to 18. 
 We sent a reading list to our participants such that they could prepare talks before the start of the course. Approximately the first half of the course was dedicated to the students' talks. Motivated by~\cite{Poelke2014Complex,Skrodzki2016Chladni}, the students asked for more math-art input, which led to discussions about fractals. Part of the second half was devoted to learning \LaTeX~as well as to writing a detailed chapter about the covered contents, see below. However, the largest time frame in the second half of the course was taken by the development of the participants' own artistic projects.

To develop their own ideas, we first held a ``gallery'' with all participants. During this activity, each student developed their own poster with a more or less concrete project idea. They were asked to pick either a handcrafted or a digital artwork to create. In a second step, all participants and the tutors examined the posters and provided further questions or suggestions via post-it notes on the posters. Each participant was free to take or reject any suggestions obtained during this step. Afterwards, the group was split. About one half worked with canvas and paint, while the other half developed digital artwork. 

\section*{Goals and Results of the Course}

The \emph{Organization for Economic Cooperation and Development} (OECD) defines \emph{mathematical literacy} as~\cite{OECD2003PISA}:
\begin{quote}
	\emph{Mathematical literacy is an individual's capacity to identify and understand the role that mathematics plays in the world, to make well-founded judgements and to use and engage with mathematics in ways that meet the needs of that individual's life as a constructive, concerned and reflective citizen.}
\end{quote}
In the context of our course, we derive our goals from this definition. Namely, the participants should build a common basis of mathematical knowledge and be able to present it. On the one hand, this is achieved by the presentations the participants gave in the context of the course. On the other hand, all participants should complete their own artistic project, should have understood the underlying mathematics, and should be able to communicate it to their possibly mathematically uneducated peers and family. In the following, we discuss different course outcomes to showcase the successful accomplishment of these goals.

\paragraph{Rotation}
One important part of the DSA is a ``rotation'' among the courses. It takes place roughly after the first half of the school. 
Students in each course are divided into groups of two or three. 
Each group gives a presentation on the contents covered in the course so far. 
Students and tutors alike attend presentations of courses other than their own.
They are encouraged to ask questions after the talk and to provide the speakers with feedback.
This leads to an audience comprised of people unfamiliar with the course material.
Thus, not only do the presentations have to cover newly learned material, but they also need to be given in a very accessible way.
This forces the participants out of their comfort zone by presenting contents they learned recently without the security of the tutors to correct mistakes.
In terms of the goals for the course, the concept of the rotation works towards the ability to communicate mathematical ideas to a lay person.


\paragraph{Participant Feedback}
 In the beginning, the group was split into those more interested in mathematics and those more interested in arts. After working in both fields, all participants were convinced of the connections between mathematics and arts. They all agreed that these two disciplines do not exclude each other. Especially those participants more interested in arts were surprised by their growing interest and abilities in solving mathematical problems. In an evaluation of the course---half a year after the school---one participant writes:
\begin{quote}
	\emph{What I liked most about the course was the combination of the topics ``mathematics and arts''. It was the first time for me to do this. As mathematics is not a strong point of mine, I was able to deduce and connect many aspects seen from the arts perspective.$^\mho$}
\end{quote}
This quote stresses the missing link between mathematics and arts in the German school system.
But it also highlights how students can benefit from a strong intuition in the arts when studying mathematics.
Additionally, this relationship holds the other way around as another participant writes:
\begin{quote}
	\emph{As we all arrived with different ideas---some being more interested in art than in mathematics and others the other way around---I particularly enjoyed how you managed to make the respective other discipline attractive for each of us. I was for example rather interested in mathematics, but in the course I also developed an enthusiasm for arts.$^\mho$}
\end{quote}

\paragraph{Artwork and Artists' Statements} 
The tangible output from this course consisted artwork created by the participants and a chapter for a commemorative collection. The artistic projects resulted in both handcrafted and digital work.

\begin{figure}
	\newlength{\anna}
	\setlength{\anna}{0.24\textwidth}
	\begin{subfigure}[t]{\anna}
		\includegraphics[width=1.\textwidth]{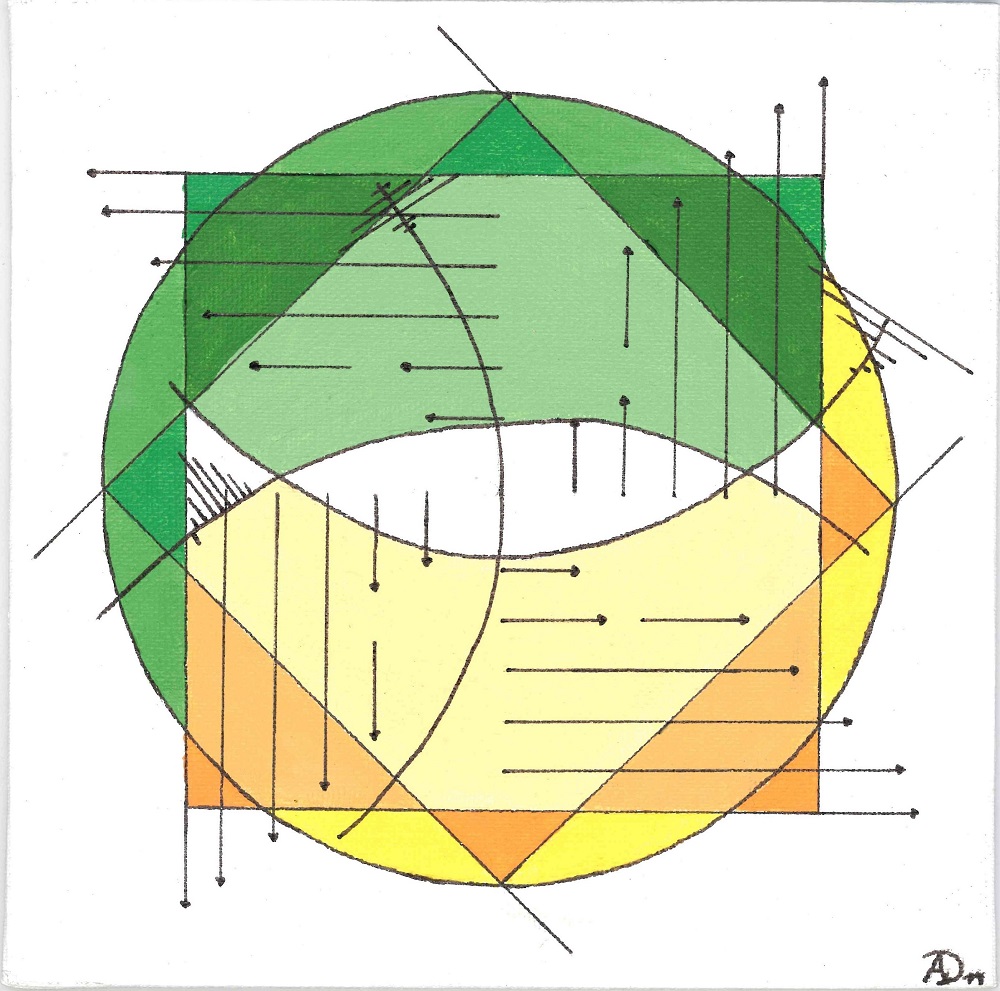}
	\end{subfigure}
	\hfill
	\begin{subfigure}[t]{\anna}
		\includegraphics[width=1.\textwidth]{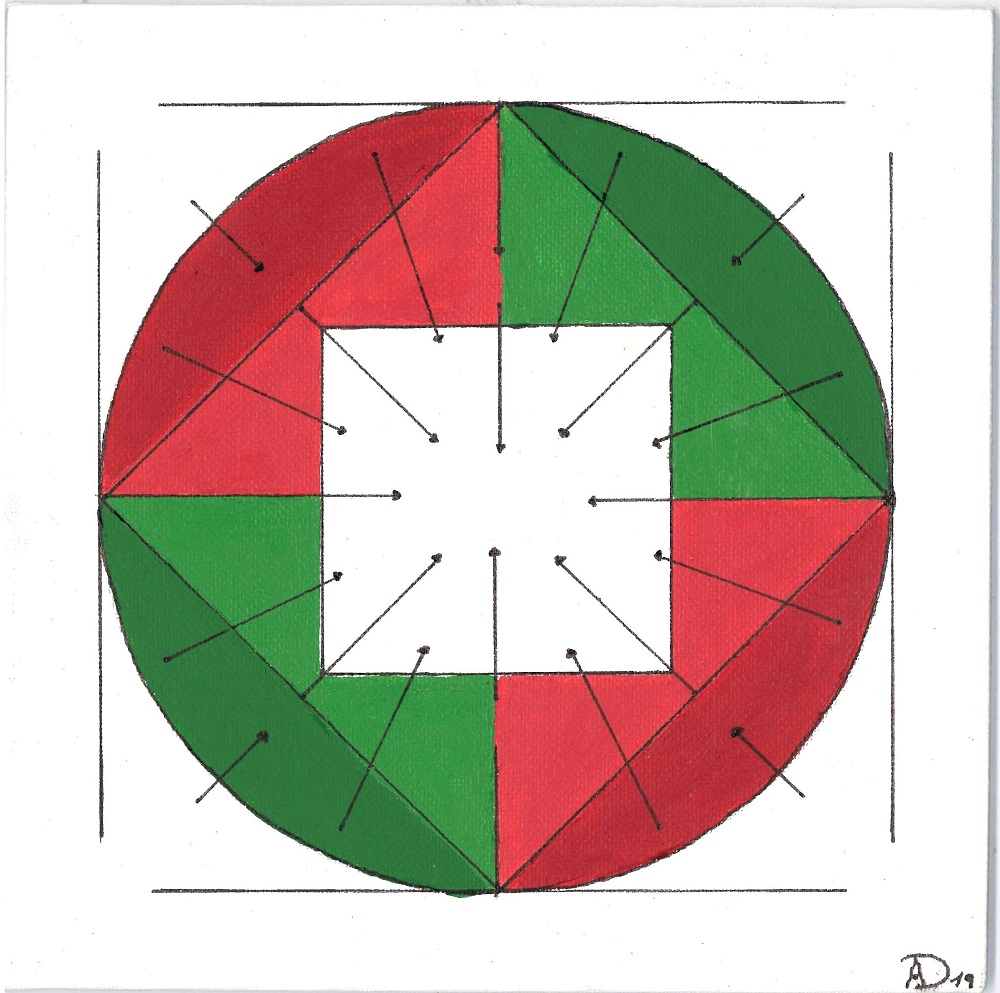}
	\end{subfigure}
	\hfill
	\begin{subfigure}[t]{\anna}
		\includegraphics[width=1.\textwidth]{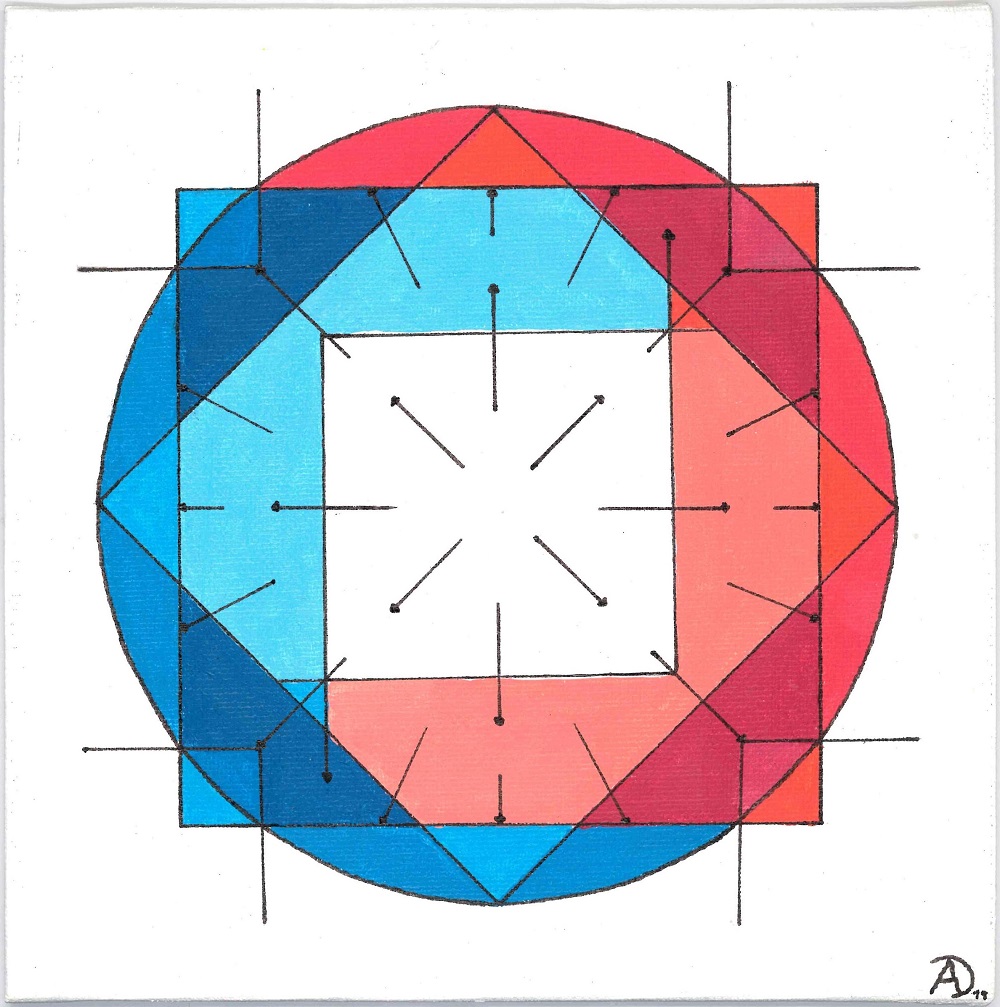}
	\end{subfigure}
	\hfill
	\begin{subfigure}[t]{\anna}
		\includegraphics[width=1.\textwidth]{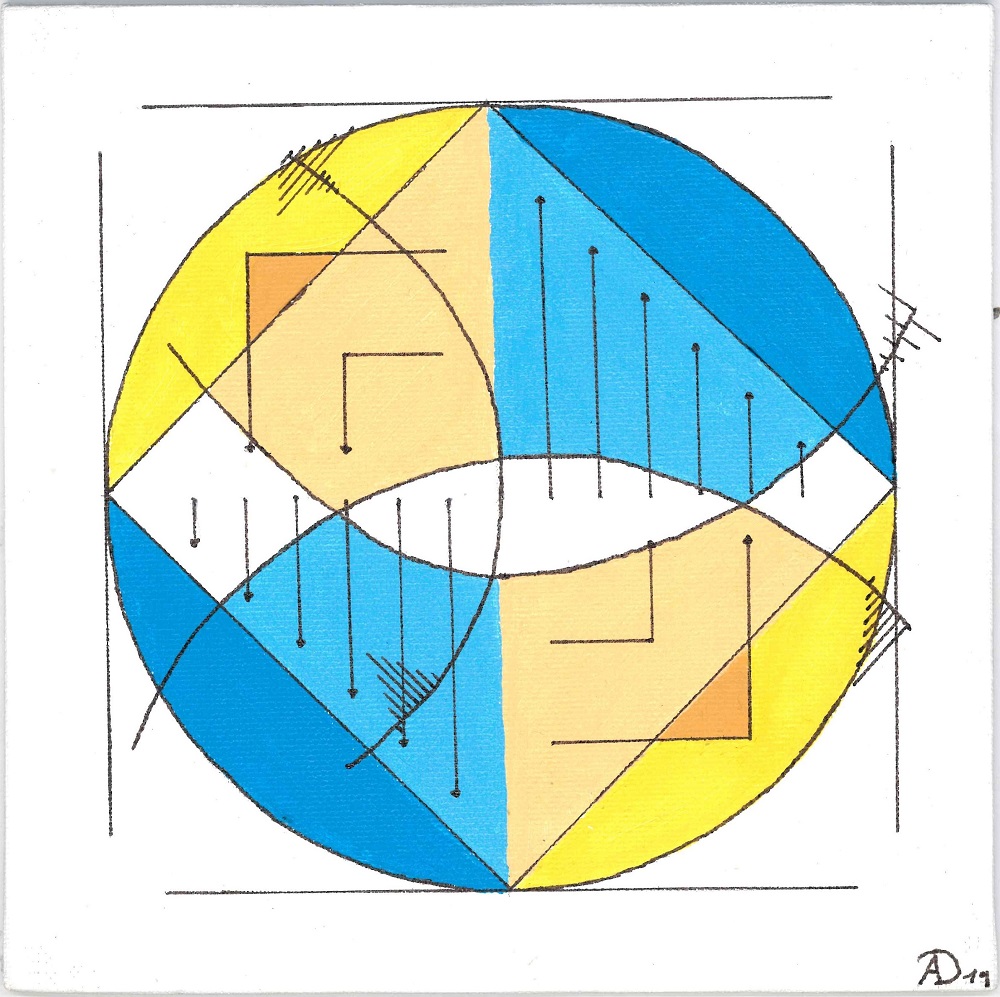}
	\end{subfigure}
	\caption{Anna Dorner, ``Ephemeros 1--4'', acrylic on canvas, 20\,cm $\times$ 20\,cm each, 2019.
	\emph{``[\kdots] The series picks up several mathematical notions such as vectors, vector fields, curl, divergence, and norms. This is accentuated through the reoccurring use of arrows as they can also be found in Paul Klee's paintings. [\kdots]
	Because of the high number of geometric forms the paintings appear to be constructed and abstract. [\kdots]
	As in [Klee's] paintings, the arrows create energy and [\kdots]~point to a goal human beings are longing for. 
	Still, humans will never reach it. [\kdots]
	[This] is intensified by rotating arrows as well as by the divergence-free vector field---everything is turning in a circle, no end is in sight. In contrast, the colorful and bright colors arouse positive feelings in the observer. The paintings appear [\kdots]
	harmonious and consistent. [\kdots]''}$^\mho$ \emph{\cite{DSA-Artwork}}
}
	\label{fig:Anna}
\end{figure}

\begin{figure}
	\newlength{\birgit}
	\setlength{\birgit}{0.3\textwidth}
	\begin{subfigure}[t]{\birgit}
		\includegraphics[width=1.\textwidth]{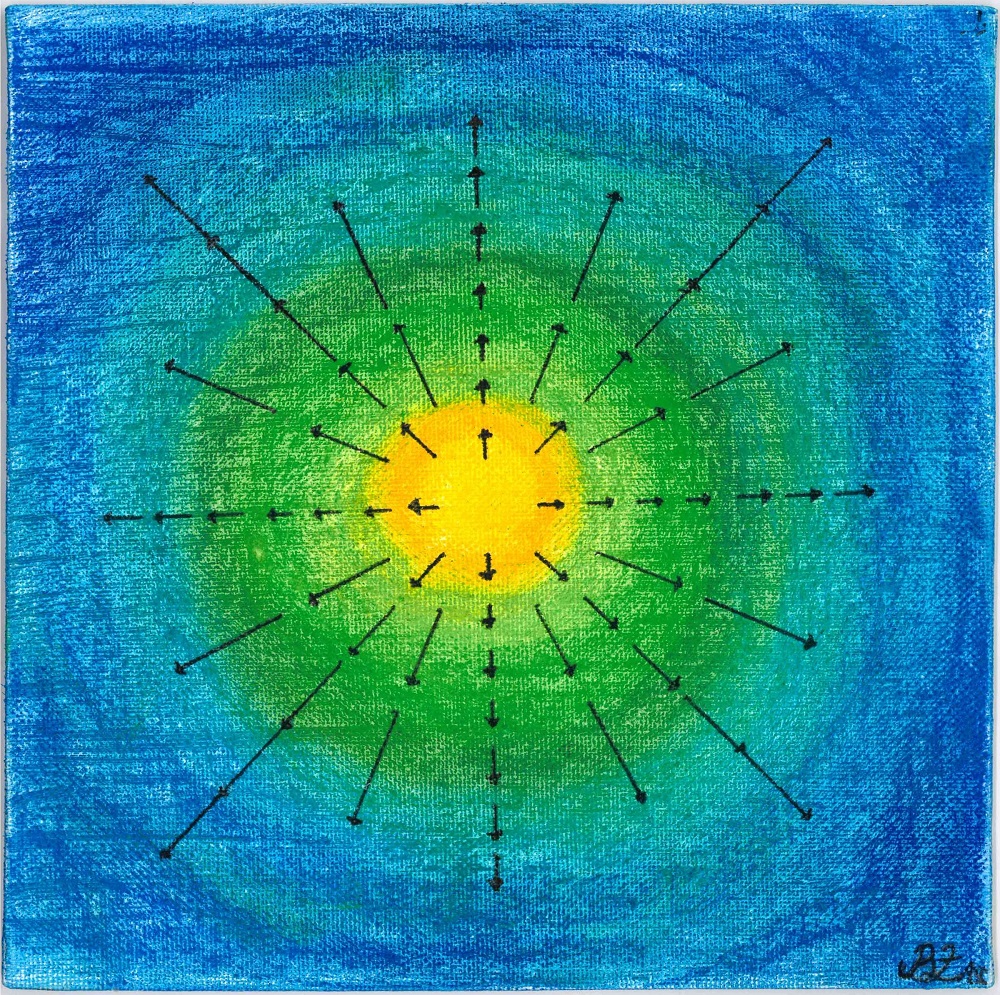}
	\end{subfigure}
	\hfill
	\begin{subfigure}[t]{\birgit}
		\includegraphics[width=1.\textwidth]{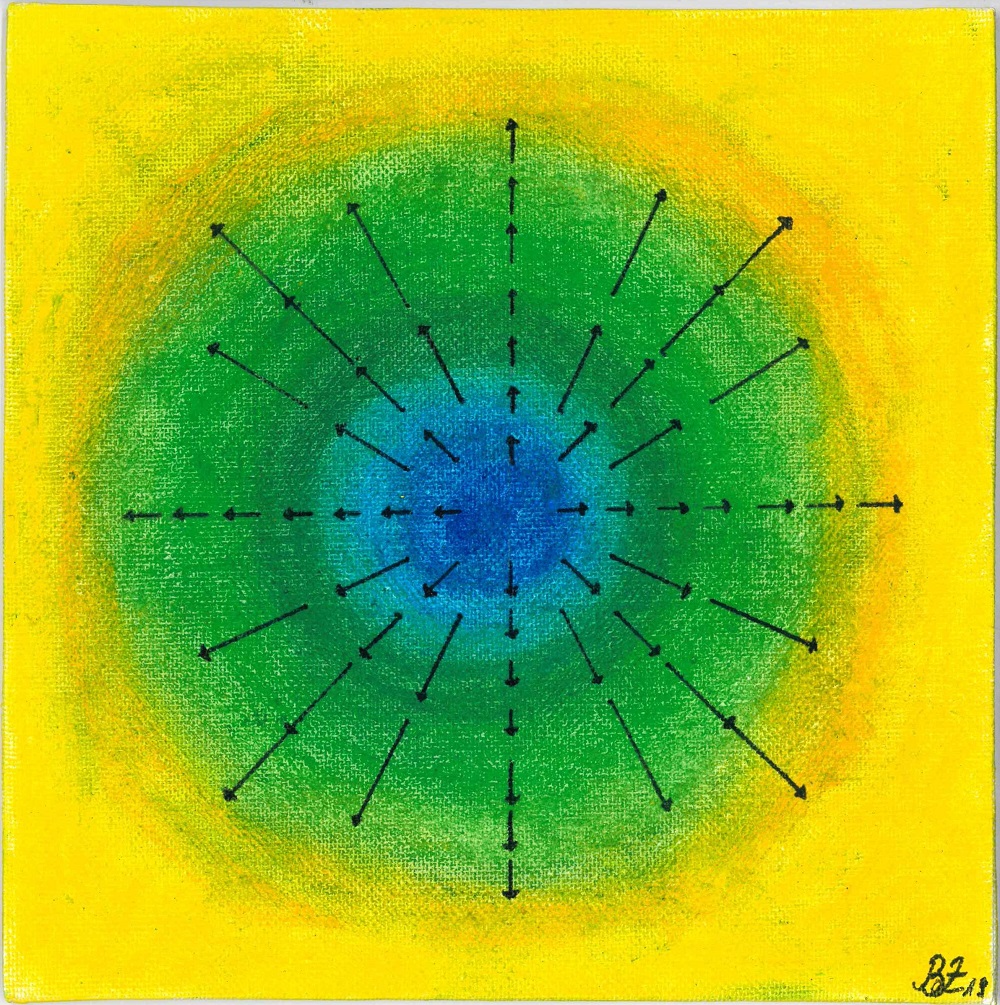}
	\end{subfigure}
	\hfill
	\begin{subfigure}[t]{\birgit}
		\includegraphics[width=1.\textwidth]{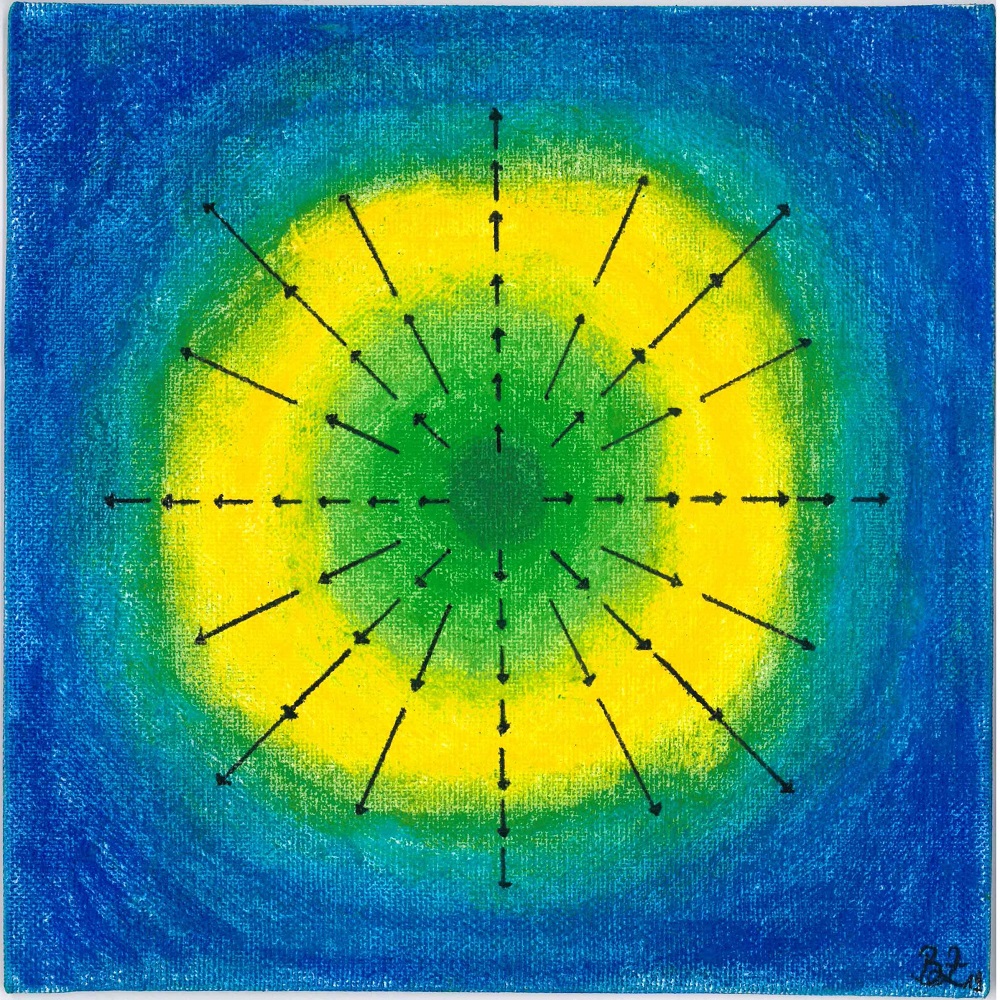}
	\end{subfigure}
	\caption{Birgit Zickler, ``Flug der Pfeile 1--3'' (flight of the arrows), pastel on canvas, 20\,cm $\times$ 20\,cm each, 2019. 
	\emph{``The motivation for ``Flight of arrows. Series in blue, green, and yellow'' is based on the idea of combining vector fields and color gradients. [\kdots] 
	Color gradients are suitable for this, since the transitions between light and dark colors can depict directions of movement.  [\kdots]
	The [generated] vector field was required to be curl-free---the vectors should radiate from the center. [\kdots]
	The [next] step consisted in the development of different color gradients which lie behind the vector field homogeneously.
	The effect to be reached consists of adding a depth effect to the two-dimensional vector field and of representing it spatially. 
	The used colors [\kdots]~were best suitable for the effect to be evoked. [\kdots]
	Pastel was chosen because it is well suited to produce homogeneous color gradients. Especially the transitions between different colors can be drawn easily. [\kdots]}$^\mho$'' \emph{\cite{DSA-Artwork}}}
	\label{fig:Birgit}
\end{figure}

\begin{figure}
	\newlength{\tessa}
	\setlength{\tessa}{0.3\textwidth}
	\begin{subfigure}[t]{\tessa}
		\includegraphics[width=1.\textwidth]{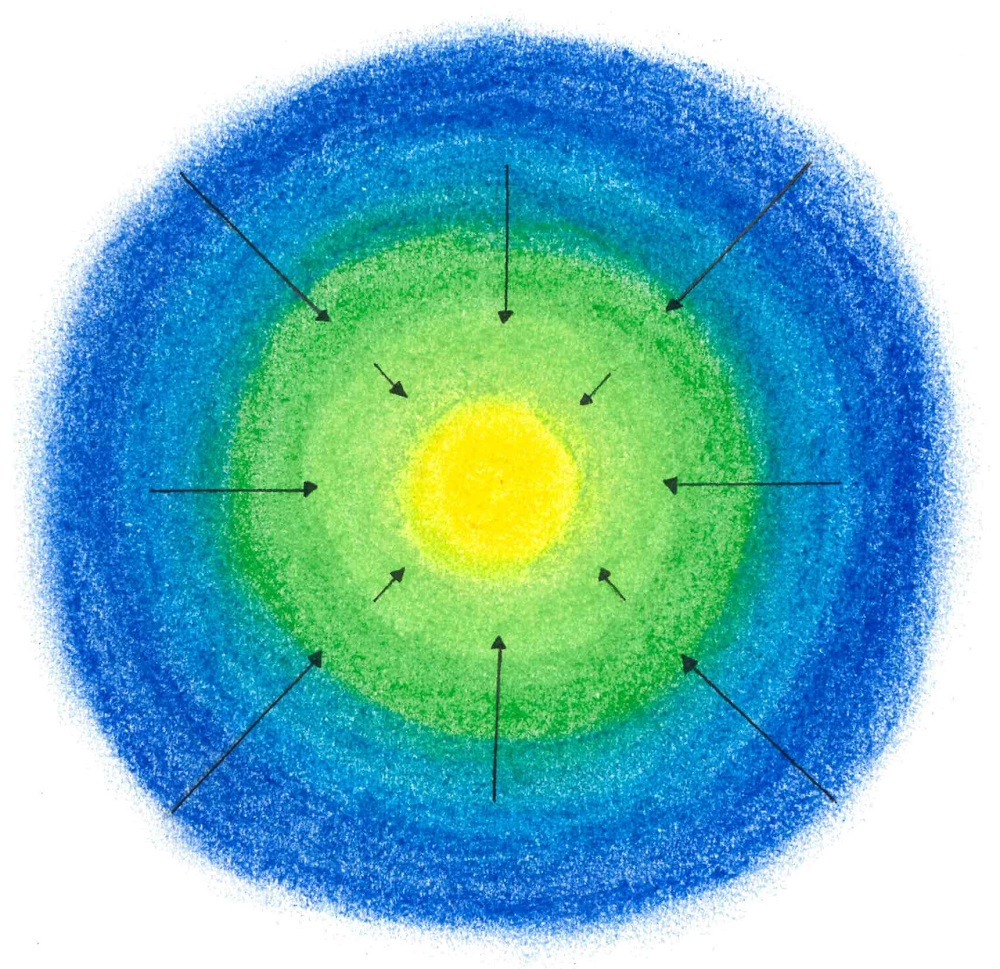}
	\end{subfigure}
	\hfill
	\begin{subfigure}[t]{\tessa}
		\includegraphics[width=1.\textwidth]{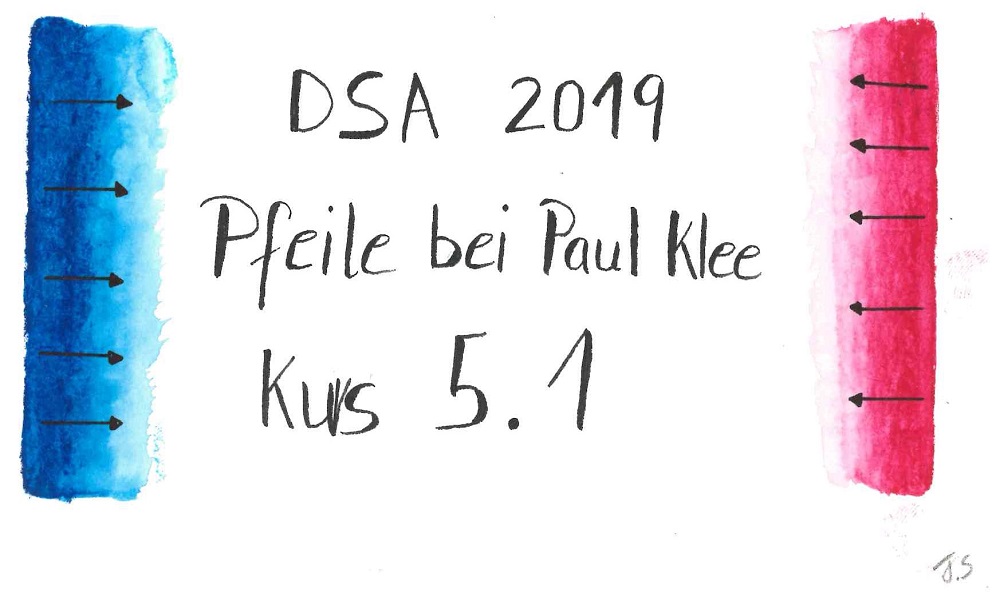}
	\end{subfigure}
	\hfill
	\begin{subfigure}[t]{\tessa}
		\includegraphics[width=1.\textwidth]{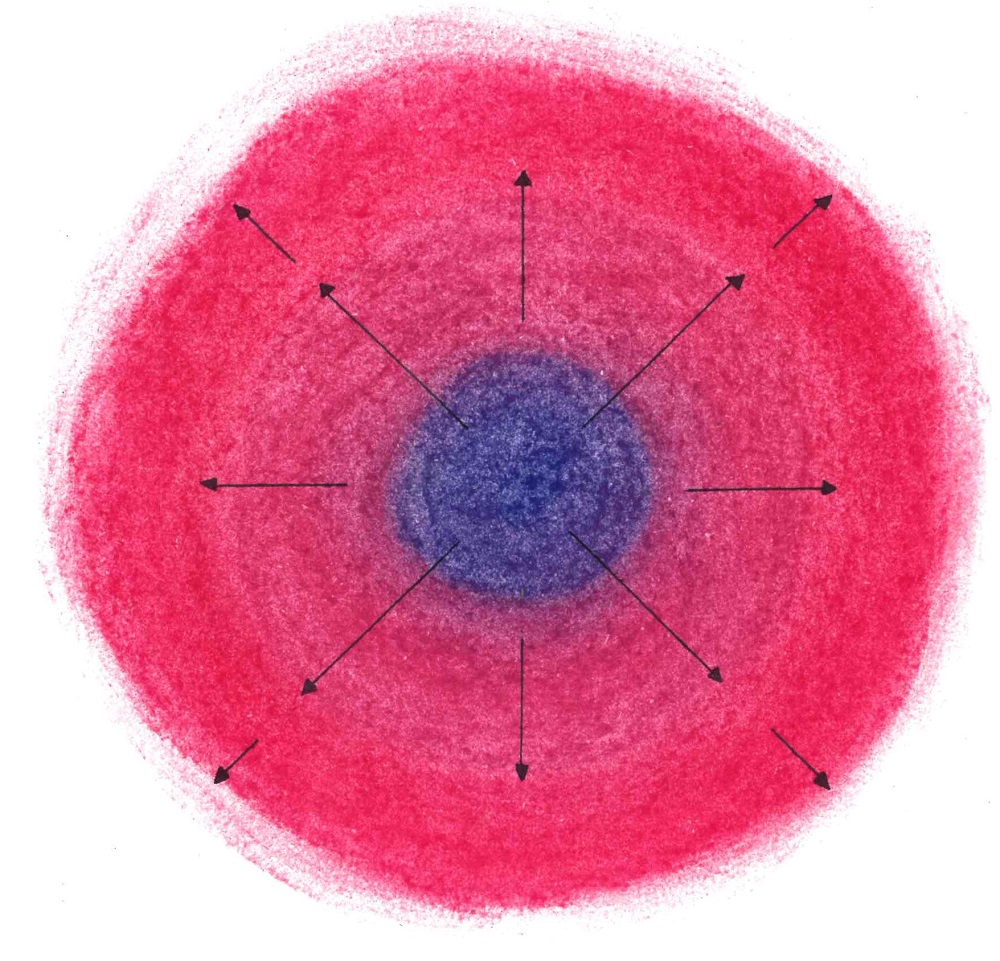}
	\end{subfigure}
	\caption{Tessa Sch\"onborn, ``Vektoren verschieben Farben 1--3'' (vectors move colors 1--3), pastel on canvas, 20\,cm $\times$ 20\,cm each, 2019. 
	\emph{
	``[\kdots] [The] idea arose to illustrate the mathematical notions divergence and vectors using colors respectively color gradients. [\kdots]
	The first and the third painting shall illustrate the notion of divergence of a fictive vector field. The first one shows a source. This is supported by the colors, which are arranged in a circle and become darker and darker towards the outside. The third painting shows a sink. Therefore, the colors become darker towards the center. The color shades are substantiated through solitary vectors. [\kdots]
	On the left hand side of the [second] painting, the blue tones of the first painting occur, on the right hand side, the red tones of the second painting occur. This represents Paul Klee's thoughts on the balance of warm and cold colors. The blue and the pink tones get brighter towards the center. Vectors support this gradient. [\kdots]}$^\mho$'' \emph{\cite{DSA-Artwork}}}
	\label{fig:Tessa}
\end{figure}

\begin{figure}
	\centering
	\newlength{\niklas}
	\setlength{\niklas}{0.3\textwidth}
	\begin{subfigure}[t]{\niklas}
		\includegraphics[width=1.\textwidth]{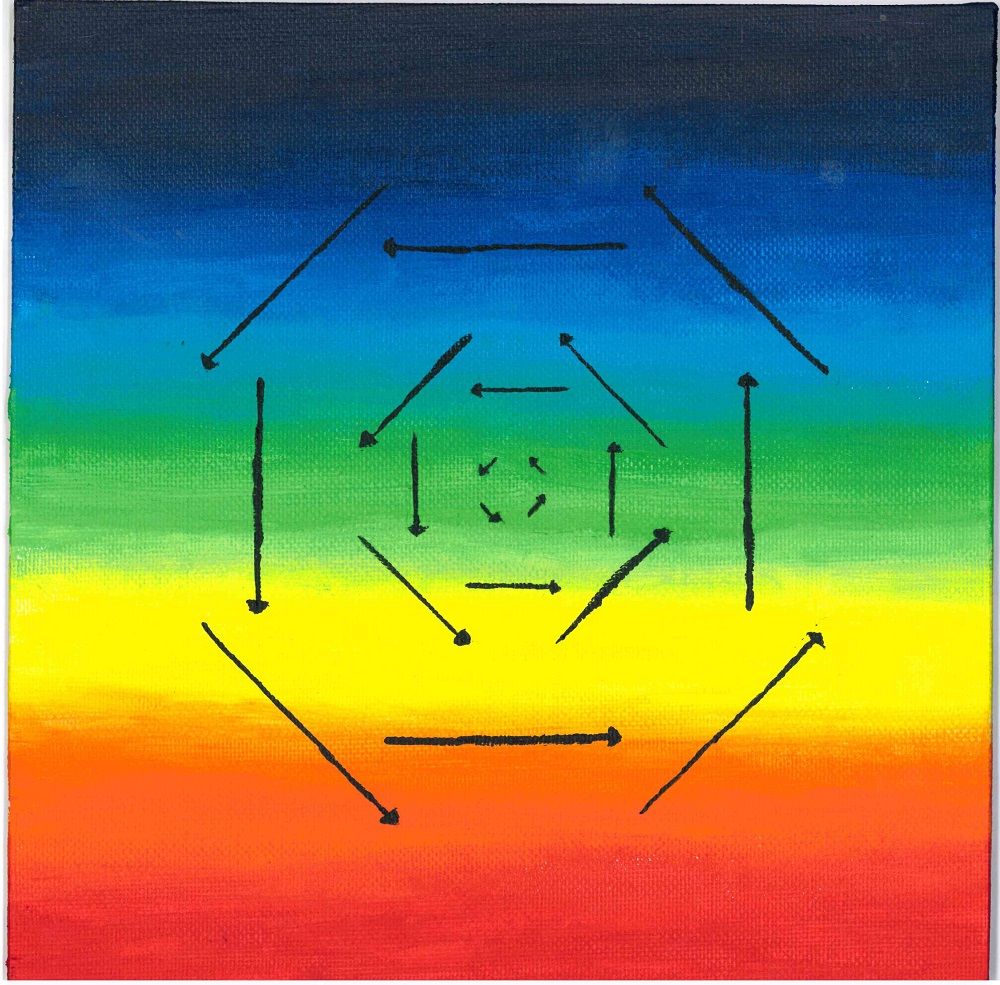}
	\end{subfigure}
	\hspace{2cm}
	\begin{subfigure}[t]{\niklas}
		\includegraphics[width=1.\textwidth]{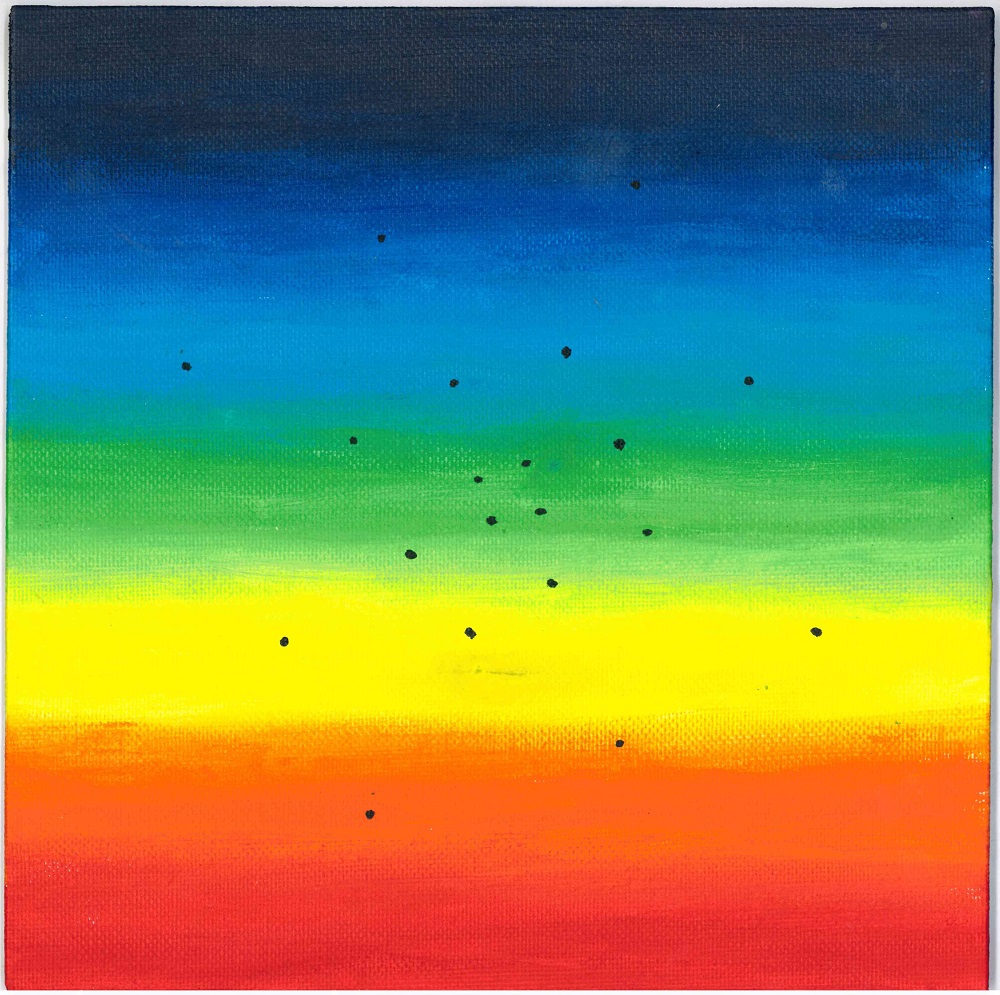}
	\end{subfigure}
	\caption{Niklas Steeger, ``Rotation 1'' and ``Rotation 2'', acrylic on canvas, 20\,cm $\times$ 20\,cm each, 2019. 
	\emph{ 
	``[\kdots] The background is designed with a color gradient in which each individual color is equally weighted. [\kdots]
	[According] to [Klee's] philosophy, every color should be equally distributed. In addition, the used colors reflect the visual range of humans and are also arranged according to their corresponding wavelength.
	The left image represents the simplified version of a rotation within a vector space. [\kdots]
	Thus, all arrows rotate around [a center] point. The further the arrows move away from the origin, the longer they are. [\kdots]
	The second image [\kdots]
	has the same gradient. [\kdots] 
	It can be seen that the points drawn in [\kdots]
	have the same coordinates as the arrowheads from the other image. Furthermore, both images can be arranged next to each other independently of the side, as long as red [\kdots] 
	is on top [or bottom], as neither the symmetry nor the gradient is destroyed by this rotation.~[\kdots]''
	}$^\mho$  \emph{\cite{DSA-Artwork}}}
	\label{fig:Niklas}
\end{figure}

\begin{figure}
	\begin{subfigure}[t]{0.31\textwidth}
		\includegraphics[width=1.\textwidth]{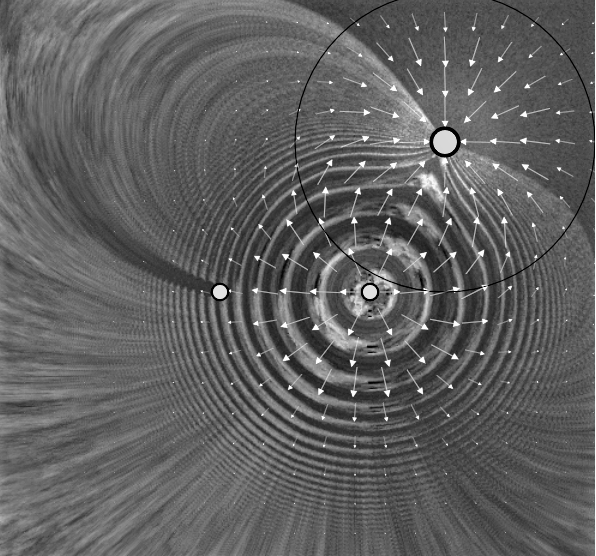}
	\end{subfigure}
	\hfill
	\begin{subfigure}[t]{0.6\textwidth}
		\includegraphics[width=1.\textwidth]{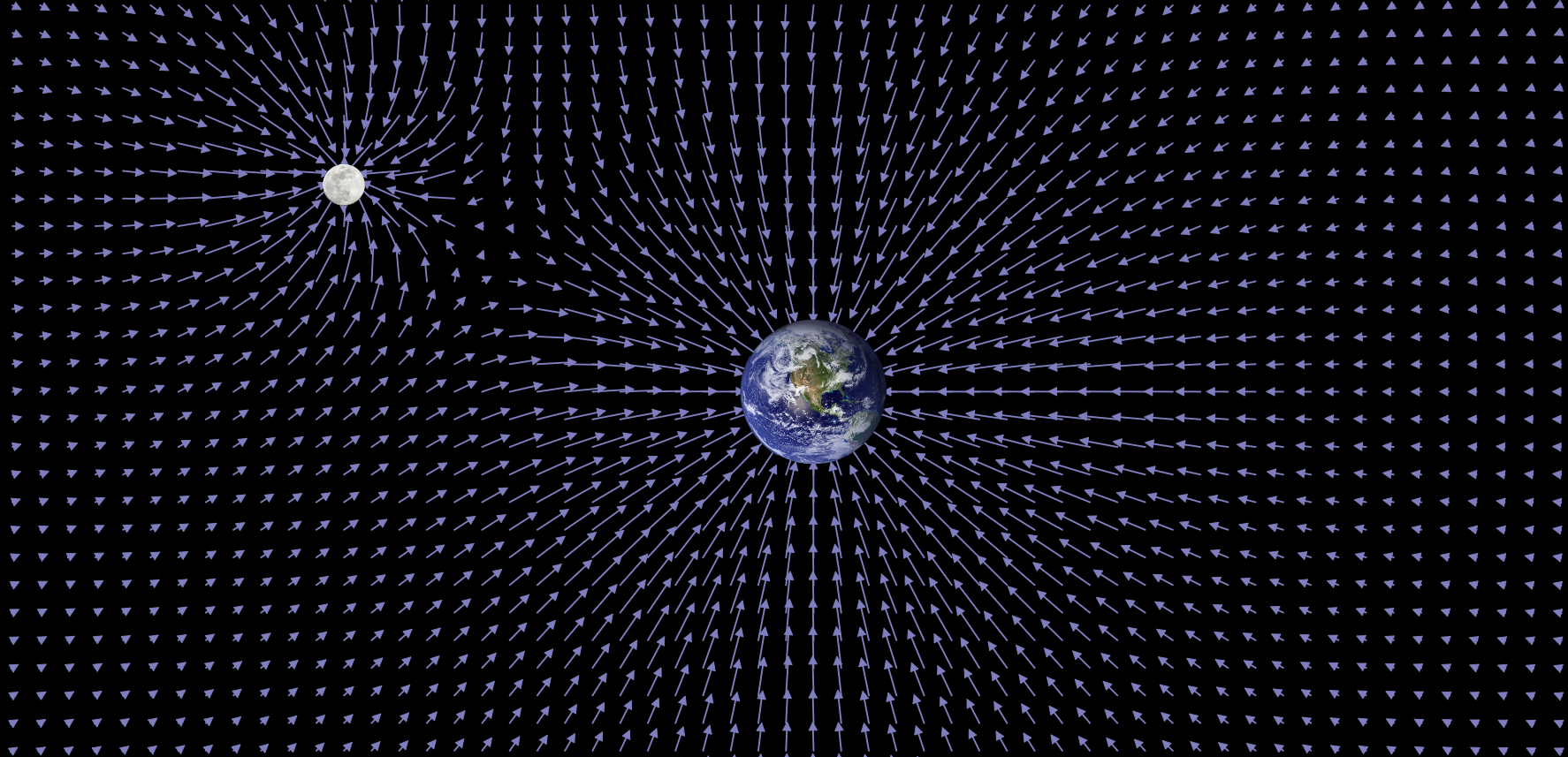}
	\end{subfigure}
	\caption{Laura Blechschmidt, Elias Dincher, Raphael Mergehenn, Lukas Schicht, and Martin Widany, ``Surrounder'', digital animation art, 2019, created using~\cite{CindyJS}. Left: Work in progress, Right: Screenshot from final animation. \emph{
			 ``The finished animation shows our planet in the center, which is regularly orbited by its satellite.
			 In the background a vector field is shown, which adapts to the circular orbit. The vector arrows show the strength and direction of the added gravitational field of the two celestial bodies. When looking at the animation more closely, it can be seen that the weight forces balance each other at some points. In addition, it can be seen that the gravitational field of the earth is stronger than that of the moon. The reason for this is that the mass of the planet is the greater, which is why it is also shown larger.''
	}$^\mho$ \emph{\cite{DSA-Artwork}}}
	\label{fig:Surrounder}
\end{figure}

The students preferring handcrafting all developed an independent and individual project. Some of these are closely related to Klee's imagination and vector fields, see Figures~\ref{fig:Anna} to~\ref{fig:Niklas}. Other artwork from the course worked with fractals, balance of colors, or color gradients.
Those participants who preferred to create digital artwork worked in two groups. One of them created an animation of the moon circling the earth, realized with CindyJS~\cite{CindyJS}. During the animation, the gravitational field created by the two objects is constantly updated. In particular the work in progress shows the underlying components and artistic choices by the participants, see Figure~\ref{fig:Surrounder}. The other group visualized a scalar field.
Additional to the rotation, the participants of our course organized an exhibition of their paintings and a presentation of their digital artwork at the last day of the school. The output received a lot of praise from all visitors.

By the end of the school, every course has to contribute a chapter about the content of their course to a commemorative collection.
While providing a memory, in writing this chapter, the participants reflect on the contents worked out and describe what they learned. While creating the artwork, we requested each participant to write a short passage about their art resp.\@ the programming projects, the mathematical contents of it, and the realization of Klee's intentions. 
The captions of Figures~\ref{fig:Anna} to~\ref{fig:Surrounder} present shortened statements by the participants from this chapter. While some work is omitted here, all artwork and the complete passages written by their creators can be found at~\cite{DSA-Artwork}. 

\section*{Discussion} 

Overall, we are very satisfied with the course and its output. Three things went particularly well and we will repeat these in a subsequent course. First, the participants each had a green, yellow, and red card in front of them for quick, non-verbal feedback. 
This method proved extremely successful and enabled the participants to easily show their current state of understanding without having to verbalize it. Second, the ``gallery'' method discussed above was very helpful in guiding the participants to provide feedback for each other. The projects all improved significantly after the execution of this activity. Third, splitting the group into a hands-on and a digital subgroup for the projects enabled the participants to follow their own interests. This way, they could use their respective skill set and approached the projects very enthusiastically.

Despite the success, we would also change several aspects of the course, if we were to repeat it. First, the level of abstraction of mathematics in the art of Paul Klee is comparably high. Other pairings---like M.C.\@ Escher and tilings or the Alhambra and planar symmetry groups---could provide an easier access for the participants to the mathematical content. The fact that some participants of our course chose to visualize fractals---which were part of the course, but unrelated to Paul Klee and vector fields---shows that the level of abstraction was probably too high for them. Second, the participants struggled severely with the presentations they were to give based on the material handed out before the course. Providing them with more targeted questions for the presentations and having them send in the presentations beforehand should help them to be more focused. Third, more time needs to be set aside to actually practice the mathematical content, i.e.\@ more time needs to be spent on exercises to become familiar with the newly learned concepts.

Finally, courses at the DSA purposely do not include any performance review of the participants. As high school students are confronted with many such tests, the absence of any review contributed tremendously to a relaxed course atmosphere. In particular, when confronting students with comparably hard contents it is important to relieve them of any additional pressure. Because of the absence of tests, we cannot make any statements on the gain of knowledge induced by the course. We take a neutral stand on the question whether or not a final exam should be included. This decision has to be made depending on the personally intended outcome of the course.

\newpage

\section*{Acknowledgments}

The DSA is organized annually by \emph{Bildung und Begabung gGmbH} and financed by the German \emph{Federal Ministry of Education and Research} (BMBF), the Stifterverband for the German Sciences, and other partners. 
The authors are thankful for the opportunity to give the presented course and thank the students for participating in the course and for their permission for printing their artwork and texts in this article.

{\setlength{\baselineskip}{13pt} 
\raggedright				

} 


\begin{thebibliography}{99}
	
	
\bibitem{DSA-Webpage}
Bildung \& Begabung gGmbH. ``Deutsche Sch\"ulerAkademie.'' \href{https://www.deutsche-schuelerakademie.de/}{www.deutsche-schuelerakademie.de}
	
\bibitem{CindyJS}
CindyJS, Version: 0.7, Technische Universit\"at M\"unchen, Germany. \url{https://cindyjs.org/}	

\bibitem{Paul2004Catalogue}
J.~Helfenstein and C.~R{\"u}melin. ``Paul Klee Catalogue Raisonn{\'e} -- Verzeichnis des gesamten Werkes in neun B{\"a}nden,'' Volumes 1--9, 2004.
	
\bibitem{Herczynski2017Paul}
A.~Herczynski. ``Paul Klee notebooks: form and mathematics.'' \emph{Talk at the Isaac Newton Institute for Mathematical Sciences}, available online at \href{https://www.newton.ac.uk/seminar/20171128141014501}{www.newton.ac.uk/seminar/20171128141014501}.
	

\bibitem{Klee1925paedagogisches}
P.~Klee. ``P{\"a}dagogisches Skizzenbuch.'' \emph{Albert Langen Verlag}, M{\"u}nchen, 1925.

\bibitem{Moesser1977pfeile}
A.~M{\"o}{\ss}er. ``Pfeile bei Paul Klee.'' \emph{Wallraf-Richartz-Jahrbuch}, vol.~39, 1977, pp.~225--235.

\bibitem{OECD2003PISA}
Organization for Economic Cooperation and Development (OECD). ``The PISA 2003 Assessment Framework -- Mathematics, Reading, Science and Problem Solving Knowledge and Skills,'' 2003. \href{https://www.oecd.org/education/school/programmeforinternationalstudentassessmentpisa/33707192.pdf}{www.oecd.org/education/school/programmeforinternationalstudentassessmentpisa/33707192.pdf}

\bibitem{Poelke2014Complex}
K.~Poelke, Z.~Tokoutsi, and K.~Polthier. ``Complex Polynomial Mandalas and their Symmetries.'' \emph{Bridges Conference Proceedings}, Seoul, Korea, Aug.~14--19, 2014, pp.~433--436. \url{https://archive.bridgesmathart.org/2014/bridges2014-433.html}

\bibitem{DSA-Artwork}
Results from ``Paul Klee and vector fields'' available at \url{https://ms-math-computer.science/dsa_19.html}

\bibitem{Skrodzki2016Chladni}
M.~Skrodzki, U.~Reitebuch, and K.~Polthier. ``Chladni Figures Revisited: A Peek Into The Third Dimension.'' \emph{Bridges Conference Proceedings}, Jyv{\"a}skyl{\"a}, Finland, Aug.~9--13, 2016, pp.~481--484. \url{https://archive.bridgesmathart.org/2016/bridges2016-481.html}
\end{thebibliography}
\end{document}